# Estimates of heterogeneity ($I^2$) can be biased in small meta-analyses

Paul T. von Hippel (University of Texas, Austin)

*Abstract*

Estimated effects vary across studies, partly because of random sampling error and partly because of heterogeneity. In meta-analysis, the fraction of variance that is due to heterogeneity is known as $I^2$. We show that the usual estimator of $I^2$ is biased. The bias is largest when a meta-analysis has few studies and little heterogeneity. For example, with 7 studies and the true value of $I^2$ at 0, the average estimate of $I^2$ is .124. Estimates of $I^2$ should be interpreted cautiously when the meta-analysis is small and the null hypothesis of homogeneity ($I^2$=0) has not been rejected. In small meta-analyses, confidence intervals may be preferable to point estimates for $I^2$.

*Supplementary materials*: A Mathematica notebook giving the author's calculations is available from the author.

*Funding acknowledgment*: This research received no specific grant from any funding agency in the public, commercial, or not-for-profit sectors.

*Correspondence*: Paul T. von Hippel, LBJ School of Public Affairs, University of Texas, 2315 Red River, Box Y, Austin, TX 78712, paulvonhippel.utaustin@gmail.com.

# 1 INTRODUCTION

When *K* different studies estimate the effect of a treatment, the estimates typically vary from one study to another. Some of this between-study variance comes from random sampling error, while some may come from *heterogeneity*. There are many sources of between-study heterogeneity, including differences in the treatment, the treated population, the study design, or the data analysis methods. It is also possible that there is no heterogeneity at all, in which case the estimates are *homogeneous* and differ only because of random sampling error.

Heterogeneity is very important. If the existing studies of a treatment are homogeneous, or only a little heterogeneous, then there is some assurance that the treatment will have a similar effect when applied to a new population. On the other hand, if the existing studies are very heterogeneous, then unless the reasons for heterogeneity are well understood, the effect of treatment on a new population will be hard to predict.

Unfortunately, when studies are compared in a meta-analysis, it is often difficult to say anything definitive about heterogeneity. The reason for this is that most meta-analyses are small; for example, half the meta-analyses in the Cochrane Library include *K*=7 studies or fewer (1). With so few studies, the classical test for heterogeneity, Cochran's *Q* (2), is not very informative because its result is as much a function of the number of studies *K* as it is of the amount of heterogeneity. If *K* is small, *Q* tends to be small and provides little power to reject the null hypothesis of homogeneity even if substantial heterogeneity is present (3). The power of *Q* and other homogeneity tests is further reduced in the unbalanced case where, for example, one of the studies in the meta-analysis is much larger than the others (3). When K is large on the other hand, Q will often reject the null hypothesis even if the amount of heterogeneity is trivial.

To better describe heterogeneity, Higgins and Thompson (4) introduced a statistic that they call $I^2$ and we call $\hat{I}_0^2$. The $\hat{I}_0^2$ statistic was meant to improve in two ways on Cochran's *Q*. First, $\hat{I}_0^2$ is more interpretable than *Q*; specifically, $\hat{I}_0^2$ estimates the proportion of the variance between study estimates that is due to heterogeneity. Second, unlike *Q*, $\hat{I}_0^2$ was meant to be independent of the number of studies *K*. Because $\hat{I}_0^2$ estimates a proportion, it ranges from 0 to 1 regardless of *K*.

$\hat{I}_0^2$ does not eliminate the uncertainty that comes from having a small number of studies in a meta-analysis. No statistic can. In small meta-analyses, for the same reason that *Q* has low power, $\hat{I}_0^2$ is very imprecise; for example, if *Q* fails to reject the null hypothesis of homogeneity, then the confidence interval around $\hat{I}_0^2$ will usually include 0. In meta-analyses from the Cochrane Reviews, the 95% confidence interval around $\hat{I}_0^2$ typically runs approximately from 0 to .60, implying that up to 60% of the between-study variance could be due to heterogeneity, or there could be no heterogeneity at all. This is not a particularly informative conclusion (1). Unfortunately, the uncertainty of the $\hat{I}_0^2$ estimate is not obvious to the typical reader of a Cochrane Review. Cochrane Reviews do not report the confidence interval around $\hat{I}_0^2$; they only report the point estimate $\hat{I}_0^2$, which may give a false sense of precision.

In this note, we show that in meta-analyses the point estimate $\hat{I}_0^2$ is not just imprecise; it is also biased. Depending on the circumstances, the bias of $\hat{I}_0^2$ can be small or large, positive or negative, but the bias is largest in the when the number of studies *K* is small and there is little true heterogeneity. For example, in a meta-analysis with $K=7$ studies and no true heterogeneity, the $\hat{I}_0^2$ statistic will on average lead us to believe that about 12% of the between-study variance is due to heterogeneity. A bias of 12% is not trivial when compared to the $\hat{I}_0^2$ values that are typically observed in meta-analyses. In meta-analyses from the Cochrane Library, for example, half of the $\hat{I}_0^2$ values are less than 21% (1).

In this remainder of this paper, we introduce background and notation and then calculate and illustrate the bias of $\hat{I}_0^2$. Having demonstrated the bias, we then discuss its implications for statistical practice.

## 2 BACKGROUND

This section introduces the notation and statistical properties that we will need to calculate the bias of $\hat{I}_0^2$.

Meta-analysis summarizes the results of $K$ studies. In each study, there is a true effect $\beta_k$ estimated by $\hat{\beta}_k$, with a true standard error $\sigma_k$ estimated by $\hat{\sigma}_k$, $k = 1, \dots, K$. Across studies, the simple average of the true effects $E(\beta_k) = \bar{\beta}$ is estimated by the precision-weighted average of the estimated effects:

$$\hat{\bar{\beta}} = \frac{\sum_{k=1}^{K} \hat{\sigma}_k^{-2} \hat{\beta}_k}{\sum_{k=1}^{K} \hat{\sigma}_k^{-2}} \qquad (1).$$

The variance of the estimated effects is partly due to the heterogeneity $\tau^2 = V(\beta_k)$ of the true effects and partly due to the standard errors $\sigma_k$:

$$\begin{aligned} V(\hat{\beta}_k) &= V(\beta_k) + V(\hat{\beta}_k - \beta_k) \\ &= \tau^2 + \sigma_k^2 \end{aligned} \qquad (2).$$

Notice that the variance $V(\hat{\beta}_k)$ is heteroskedastic because each study has a different standard error $\sigma_k$. To clearly define fractions of variance, Higgins and Thompson (4) first made the simplifying assumption that the standard errors are all equal—i.e., $\sigma_k = \sigma$. Then

$$V(\hat{\beta}_k) = \tau^2 + \sigma^2 \qquad (3),$$

and the fraction of variance that is due to heterogeneity is

$$I^2 = \frac{\tau^2}{\tau^2 + \sigma^2} \qquad (4).$$

$I^2$ is defined a little differently if the standard errors $\sigma_k$ are unequal, but we will focus here on the simple situation with equal standard errors.

The null hypothesis of homogeneity ($H_0: I^2 = 0$) can be tested by Cochran's $Q$ statistic (2):

$$Q = \sum_{k=1}^{K} \frac{\left(\hat{\beta}_k - \hat{\bar{\beta}}\right)^2}{\hat{\sigma}_k^2} \quad (5).$$

Under $H_0$, $Q$ has approximately a central chi-square distribution with $df = K - 1$ degrees of freedom. By convention, the null hypothesis is rejected if the chi-square test has $p<.1$ or $p<.05$.

Under the alternative hypothesis of heterogeneity ($H_1: I^2 > 0$), $Q$ has asymptotically a non-central chi-square distribution with $df$ degrees of freedom and a non-centrality parameter of (5)

$$\lambda = \sum_{k=1}^{K} \frac{(\beta_k - \bar{\beta})^2}{\sigma_k^2} \quad (6).$$

If again we make the simplifying assumption $\sigma_k = \sigma$ then the non-centrality parameter reduces to

$$\begin{aligned}\lambda &= \frac{1}{\sigma^2}\sum_{k=1}^{K}(\beta_k - \bar{\beta})^2 \\ &= K\frac{\tau^2}{\sigma^2} \\ &= K\frac{I^2}{1-I^2}\end{aligned} \quad (7).$$

The last line is a useful expression because it shows that $\lambda$ is an increasing function of $I^2$. The last line also has the intuitive implication that $\lambda = 0$ if $I^2 = 0$; in other words, as the fraction of variance due to heterogeneity gets small, $Q$ converges toward the central chi-square distribution that it has under homogeneity.

## 3 BIAS OF THE ESTIMATOR $\hat{I}_0^2$

To estimate $I^2$, Higgins and Thompson (4) first derived the estimator

$$\hat{I}^2 = 1 - \frac{df}{Q} \quad (8).$$

They noticed, however, that $\hat{I}^2$ can be negative though the estimand $I^2$ cannot. Negative values of $\hat{I}^2$ are not rare. Under $H_0$, when $Q$ has a central chi-square distribution, the probability of negative values exceeds 50% because $\hat{I}^2$ is negative whenever $Q<df$.

To avoid negative estimates, Higgins and Thompson (4) suggested rounding them up to zero. The rounded estimator

$$\hat{I}_0^2 = \max(0, \hat{I}^2) \tag{9}.$$

is the estimator that is most widely used today.

In the following sections, we calculate the bias of $\hat{I}_0^2$. We start by calculating the bias under the null hypothesis of homogeneity, and then extend the calculations to cover the alternative hypothesis of heterogeneity.

### 3.1 Under homogeneity

Under the null hypothesis of homogeneity ($H_0: I^2 = 0$), $\hat{I}_0^2$ has a positive bias. The reason for the positive bias can be described intuitively. $\hat{I}_0^2$ can only take values that are positive or zero, and the average of those values is necessarily positive, which means that the expectation of $\hat{I}_0^2$ is positive and exceeds the estimand of $I^2 = 0$.

Calculating the size of the bias requires a little more effort. The probability that $\hat{I}_0^2 = 0$ is $P(Q < df)$, and the probability that $\hat{I}_0^2 > 0$ is $P(Q > df)$. Therefore the expectation of $\hat{I}_0^2$ is

$$\begin{aligned} E(\hat{I}_0^2) &= P(Q < df) \times 0 + P(Q > df) \times E(\hat{I}_0^2 | Q > df) \\ &= P(Q > df) \times E\left(1 - \frac{df}{Q} \Big| Q > df\right) \end{aligned} \tag{10}.$$

Because $Q$ has a central chi-square distribution, the expression in (10) has a closed-form solution

$$E(\hat{I}_0^2) = \left(\frac{df}{df-2}\right) \frac{\left(\frac{df}{2e}\right)^{df/2} - \Gamma\left(\frac{df}{2}, \frac{df}{2}\right)}{\Gamma\left(\frac{df}{2} + 1\right)} \tag{11},$$

which we obtained using Mathematica software, version 8. In the denominator $\Gamma\left(\frac{df}{2} + 1\right)$ is the gamma function, which has one argument. In the numerator, $\Gamma\left(\frac{df}{2}, \frac{df}{2}\right)$ is the upper incomplete gamma function, which has two arguments.

Figure 1 plots $E(\hat{I}_0^2)$ as a function of the number of studies $K=df+1$. Since $I^2 = 0$, the expectation of $\hat{I}_0^2$ is also the bias. The bias is positive, and shrinks at a decreasing rate as $K$ grows. With $K=5$ studies the bias is .135; with $K=10$ studies the bias is .11; With $K=50$ studies the bias is .06.

### 3.2 Under heterogeneity

Under the alternative hypothesis of heterogeneity, $I^2 > 0$ and $Q$ has a noncentral chi-square distribution with *df* degrees of freedom and a noncentrality parameter of $\lambda = KI^2/(1 - I^2)$ according to equation (7). The expectation $E(\hat{I}_0^2)$ is still given by equation (10) but no longer reduces to expression (11) or any other closed-form expression. Instead, to evaluate $E(\hat{I}_0^2)$ we use numerical integration in Mathematica.

Figure 2 plots $E(\hat{I}_0^2)$ as a function of the number of studies $K$ when $I^2 = .05$. A dotted line is drawn at .05, so that the bias of $\hat{I}_0^2$ is the difference between the dotted line and the curve $E(\hat{I}_0^2)$. The shape of Figure 2 is very similar to the shape of Figure 1, and the bias in Figure 2 is just slightly smaller than the bias in Figure 1. What this means is that, when the amount of heterogeneity is very small, the bias is very similar to the bias under homogeneity.

Figure 3 is a graphics grid displaying 9 plots of $E(\hat{I}_0^2)$ as a function of $K$ for values of $I^2$ between .1 and .9. The bias is generally larger for small $K$. At $I^2 = .1$ the bias is positive but smaller than the bias at $I^2 = .05$ or $I^2 = 0$. At $I^2 = .2$ there is practically no bias except for very small $K$. Above $I^2 = .2$ the bias switches from positive to negative. As $I^2$ increases from .3 to .5 the negative bias gets larger, but as $I^2$ increases further from .6 to .7, the bias gets smaller and is increasingly restricted to small values of $K$, until at $I^2=.8$ there is practically no bias. At $I^2=.9$ the bias is positive again but very small and restricted to the very small values of $K$.

## DISCUSSION

We have shown that, in small meta-analyses, the widely used heterogeneity statistic $\hat{I}_0^2$, which was already known to be imprecise, is biased as well. The bias is negligible around $I^2 = .2$ or $I^2 = .8$ but worse around $I^2 = .5$ and worst when $I^2$ is close to 0.

The bias and imprecision of $\hat{I}_0^2$ are unavoidable and should not be taken as a criticism of the statistic itself. All statistics are imprecise in small samples, and any reasonable estimator of $I^2$ will be biased when $I^2$ is close to 0. The reason for the bias is fundamental. Any reasonable estimator will be limited to values of 0 or greater, so the average of these values will be positive and will necessarily exceed the estimand $I^2$ when the true value of $I^2$ is close to 0. Bias cannot be altogether avoided.

Despite its bias and imprecision, the $\hat{I}_0^2$ statistic remains useful. In large meta-analyses, $\hat{I}_0^2$ can be precise with little bias, and even in small meta-analyses it is better to have some estimate of $I^2$ than it is to have no estimate at all. In addition, although the bias of $\hat{I}_0^2$ depends on the number of studies $K$, $\hat{I}_0^2$ is much less dependent on $K$ than $Q$ is. Nevertheless, $\hat{I}_0^2$ should be interpreted very cautiously in small meta-analyses, especially when the null hypothesis of homogeneity ($I^2 = 0$) has not been rejected. For if the null hypothesis is true, or almost true, then $\hat{I}_0^2$ will be positively biased.

Perhaps the most straightforward response to the bias and imprecision of $\hat{I}_0^2$ is to report a 95% confidence interval for $I^2$ in addition to—or even instead of—the point estimate $\hat{I}_0^2$. Although the best formulas for calculating $I^2$ confidence intervals are a bit complicated (4,5), they have good coverage and they give a sense of the range of possible $I^2$ values without highlighting a point estimate that may be biased and imprecise. Some meta-analyses do report confidence intervals for $I^2$ (6), but the Cochrane Reviews do not. The Cochrane Collaboration should consider changing this practice.

In small meta-analyses, confidence intervals for $I^2$ are often very wide (1) but their width tells us something. The width of the confidence intervals tells us how little information a small meta-analysis typically provides about heterogeneity. In many small meta-analyses, we may not be able to estimate heterogeneity with much precision; in fact, we may have little confidence in any estimate except for the average effect size. No statistic can change the limitations of small meta-analyses, and the statistics that we report should make those limitations clear.

**FIGURES**

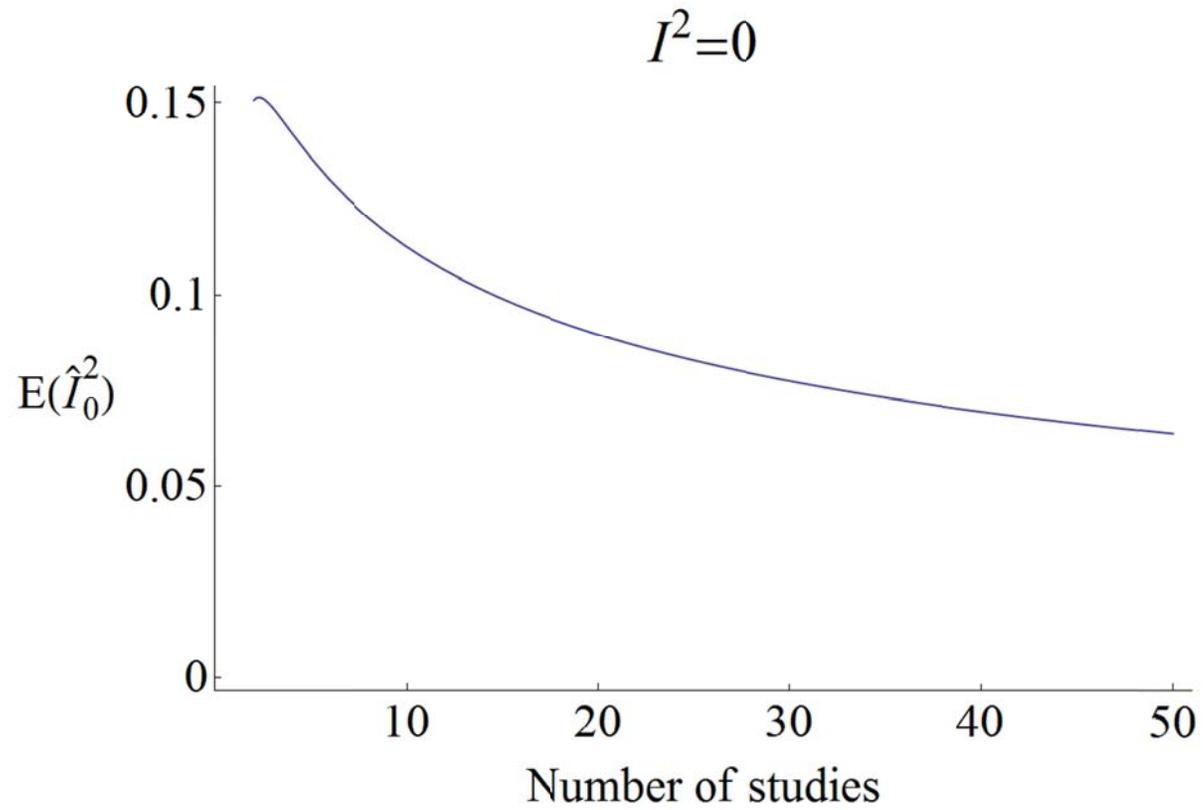

Figure 1. The expectation of $\hat{I}_0^2$ when $I^2 = 0$.

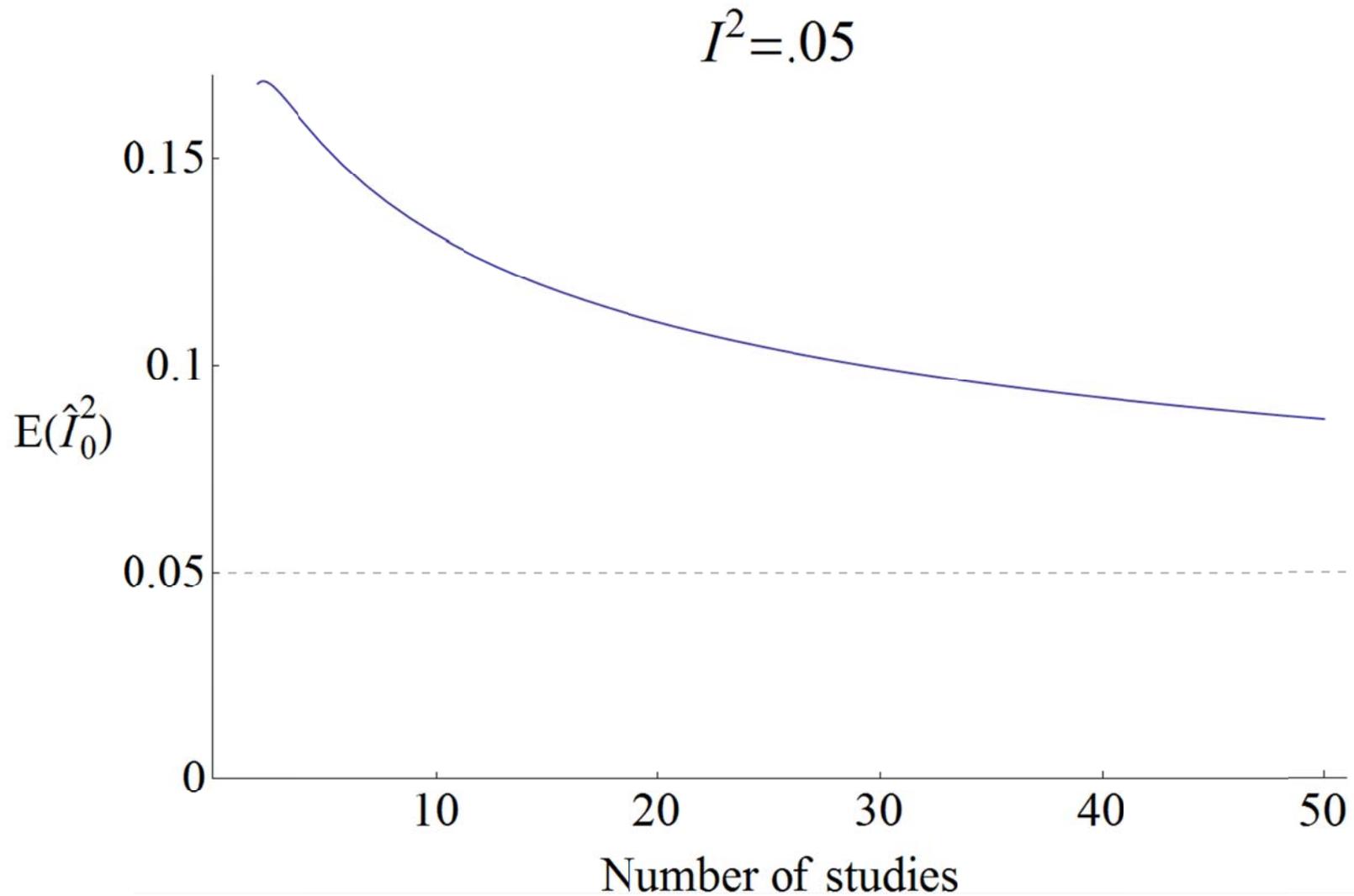

Figure 2. The expectation of $\hat{I}_0^2$ when $I^2 = .05$.

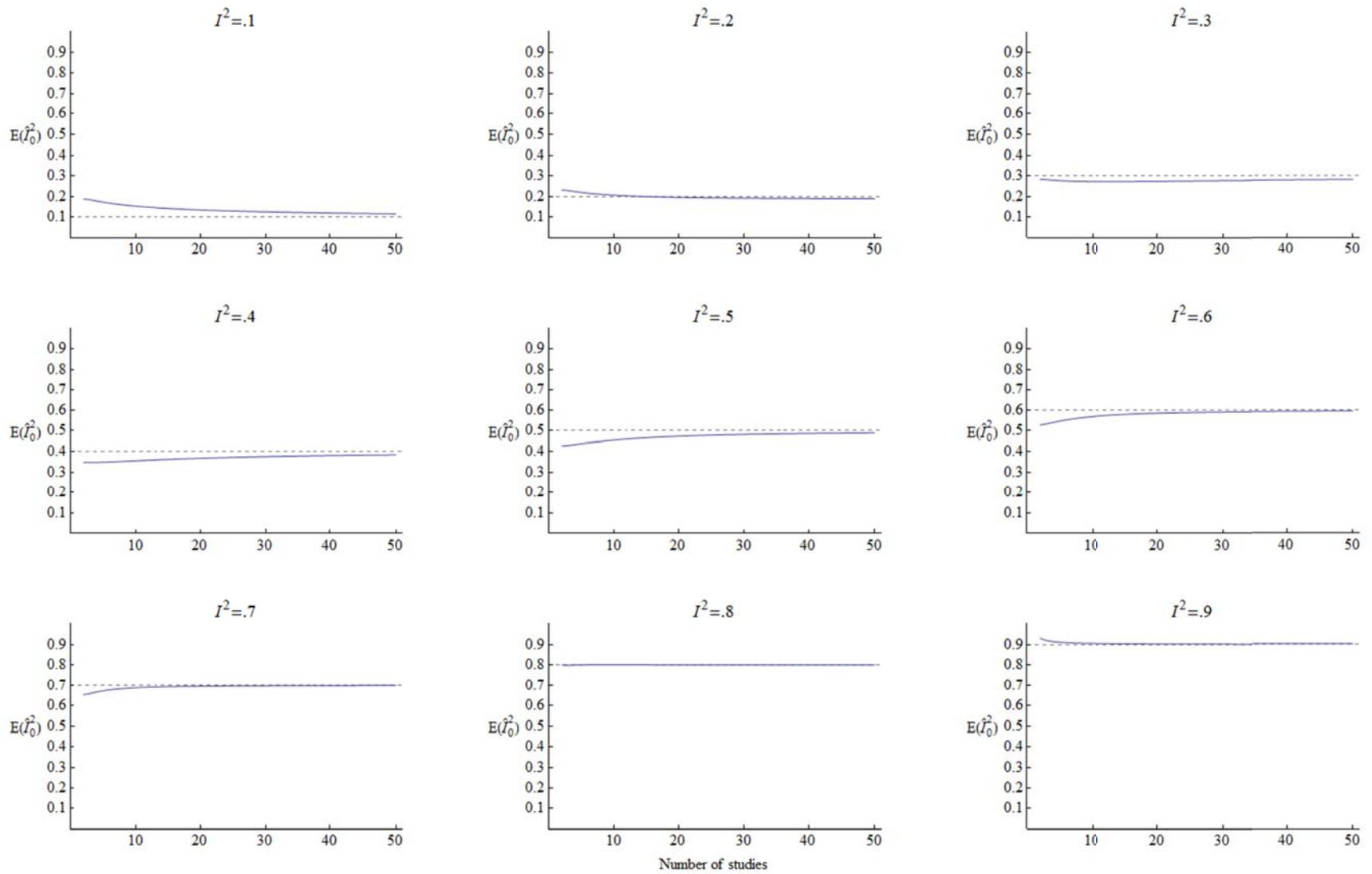

Figure 3. The expectation of $\hat{I}_0^2$ when $I^2 = .1, .2, \ldots, .9$.